\documentclass[aps,prb,a4paper,10pt,twocolumn,showpacs,floatfix,superscriptaddress,amsmath,amsfonts,amssymb,preprintnumbers,longbibliography]{revtex4-1}
\usepackage{float}

\usepackage{dcolumn,graphicx,color,booktabs,microtype,afterpage}
\graphicspath{{./}{figure/}}
\usepackage[charter,greekuppercase=italicized]{mathdesign}
\usepackage{sidecap}
\renewcommand{\tablename}{Table}
\makeatletter\renewcommand{\fnum@figure}[1]{\figurename~\thefigure.~}\makeatother
\makeatletter\renewcommand{\fnum@table}[1]{\tablename~\thetable.}\makeatother

\newcount\hh \newcount\mm
\hh=\time \divide\hh by 60
\mm=\hh \multiply\mm by 60 \mm=-\mm
\advance\mm by \time
\def\now{\number\hh:\ifnum\mm<10{}0\fi\number\mm}

\usepackage[colorlinks,plainpages=false,linkcolor=blue,urlcolor=blue,citecolor=blue,pdfpagemode=UseNone,pdfstartview=FitBH]{hyperref}

\begin{document}

\makeatletter\renewcommand{\ps@plain}{%
\def\@evenhead{\hfill\itshape\rightmark}%
\def\@oddhead{\itshape\leftmark\hfill}%
\renewcommand{\@evenfoot}{\hfill\small{--~\thepage~--}\hfill}%
\renewcommand{\@oddfoot}{\hfill\small{--~\thepage~--}\hfill}%
}\makeatother\pagestyle{plain}

\preprint{\textit{Preprint: \today, \now.}} 

%
%
\title{Nodeless superconductivity and time-reversal symmetry breaking in the noncentrosymmetric superconductor Re$_{24}$Ti$_{5}$}
\author{T.\ Shang}\email[Corresponding authors:\\]{tian.shang@psi.ch}
\affiliation{Laboratory for Multiscale Materials Experiments, Paul Scherrer Institut, Villigen CH-5232, Switzerland}
\affiliation{Swiss Light Source, Paul Scherrer Institut, Villigen CH-5232, Switzerland}
\affiliation{Institute of Condensed Matter Physics, \'Ecole Polytechnique F\'ed\'erale de Lausanne (EPFL), Lausanne CH-1015, Switzerland.}
\author{G.\ M.\ Pang}
\affiliation{Department of Physics, Zhejiang University, Hangzhou, 310058, China}
\affiliation{Center for Correlated Matter, Zhejiang University, Hangzhou, 310058, China}
\author{C.\ Baines}
\affiliation{Laboratory for Muon-Spin Spectroscopy, Paul Scherrer Institut,
CH-5232 Villigen PSI, Switzerland}
\author{W.\ B.\ Jiang}
\affiliation{Department of Physics, Zhejiang University, Hangzhou, 310058, China}
\affiliation{Center for Correlated Matter, Zhejiang University, Hangzhou, 310058, China}
\author{W.\ Xie}
\affiliation{Department of Physics, Zhejiang University, Hangzhou, 310058, China}
\affiliation{Center for Correlated Matter, Zhejiang University, Hangzhou, 310058, China}
\author{A.\ Wang}
\affiliation{Department of Physics, Zhejiang University, Hangzhou, 310058, China}
\affiliation{Center for Correlated Matter, Zhejiang University, Hangzhou, 310058, China}
\author{M.\ Medarde}
\affiliation{Laboratory for Multiscale Materials Experiments, Paul Scherrer Institut, Villigen CH-5232, Switzerland}
\author{E.\ Pomjakushina}
\affiliation{Laboratory for Multiscale Materials Experiments, Paul Scherrer Institut, Villigen CH-5232, Switzerland}
\author{M.\ Shi}
\affiliation{Swiss Light Source, Paul Scherrer Institut, Villigen CH-5232, Switzerland}
\author{J.\ Mesot}
\affiliation{Paul Scherrer Institut, CH-5232 Villigen PSI, Switzerland}
\affiliation{Institute of Condensed Matter Physics, \'Ecole Polytechnique F\'ed\'erale de Lausanne (EPFL), Lausanne CH-1015, Switzerland.}
\affiliation{Laboratorium f\"ur Festk\"orperphysik, ETH Z\"urich, CH-8093 Zurich, Switzerland}
\author{H.\ Q.\ Yuan}\email{hqyuan@zju.edu.cn}
\affiliation{Department of Physics, Zhejiang University, Hangzhou, 310058, China}
\affiliation{Center for Correlated Matter, Zhejiang University, Hangzhou, 310058, China}
\author{T.\ Shiroka}\email{tshiroka@phys.ethz.ch}
\affiliation{Laboratorium f\"ur Festk\"orperphysik, ETH Z\"urich, CH-8093 Zurich, Switzerland}
\affiliation{Paul Scherrer Institut, CH-5232 Villigen PSI, Switzerland}

\begin{abstract}
The noncentrosymmetric superconductor Re$_{24}$Ti$_{5}$, a time-reversal 
symmetry (TRS) breaking candidate with $T_c = 6$\,K, was studied by means of 
muon-spin rotation/relaxation ($\mu$SR) and tunnel-diode oscillator (TDO) techniques. At a macroscopic
level, its bulk superconductivity was investigated via electrical resistivity, magnetic
susceptibility, and heat capacity measurements. The low-temperature penetration depth, superfluid 
density and electronic heat capacity all evidence an $s$-wave coupling with an enhanced superconducting gap. The spontaneous
magnetic fields revealed by zero-field $\mu$SR below $T_c$ indicate
a time-reversal symmetry breaking and thus the unconventional nature
of superconductivity in Re$_{24}$Ti$_{5}$. The concomitant occurrence of 
TRS breaking also in the isostructural Re$_6$(Zr,Hf) compounds, hints at its common 
origin in this superconducting family and that an enhanced spin-orbital coupling does not affect pairing symmetry. 
\end{abstract}



\maketitle\enlargethispage{3pt}

\vspace{-5pt}
%
Superconductors with an inversion center can host either even-parity spin-singlet or odd-parity spin-triplet states. These strict symmetry-imposed requirements, however, are relaxed in noncentrosymmetric superconductors (NCSCs), where parity-mixed superconducting states are also allowed.
In these materials the lack of an inversion symmetry often induces an antisymmetric spin-orbit coupling (ASOC), which can lift the degeneracy of conduction band electrons.  Since the extent of parity-mixing is determined by the strength of SOC, formally similar compounds, but with different
spin-orbit couplings, can exhibit different degrees of parity mixing.

The recent interest in NCSCs is related to the complex nature of their superconducting properties.\cite{Bauer2012,smidman2017} Because of the mixed pairing,  noncentrosymmetric superconductors can display significantly different properties compared to their conventional counterparts. Some NCSCs,  such as CePt$_3$Si,\cite{bonalde2005CePt3Si},CeIrSi$_3$,\cite{mukuda2008CeIrSi3}, Li$_2$Pt$_3$B,\cite{yuan2006,nishiyama2007} and Mo$_3$Al$_2$C,\cite{bauer2010} exhibit line nodes, while others, as LaNiC$_2$,\cite{chen2013} and (La,Y)$_2$C$_3$,\cite{kuroiwa2008} show multiple superconducting gaps.
Furthermore, because of the spin-triplet pairing, the upper critical field often exceeds the Pauli limit, as has been found, e.g., in CePt$_3$Si,\cite{bauer2004} and Ce(Rh,Ir)Si$_3$.\cite{kimura2007,sugitani2006} Finally, some NCSCs, as e.g., LaNiC$_2$,\cite{Hillier2009}, Re$_6$(Zr,Hf)\cite{Singh2014,Singh2017} and La$_7$Ir$_3$,\cite{Barker2015} are known to break the time-reversal symmetry (TRS).

The binary alloy Re$_{24}$Ti$_{5}$ is a NCSC with superconducting temperature $T_c = 6$\,K, as reported already in the 1960s'.\cite{Matthias1961}
Its physical properties were studied in detail only recently,\cite{Lue2013ReTi} yet to date the microscopic nature of its SC remains largely unexplored. Similarly to Re$_{24}$Zr$_{5}$ and Re$_{24}$Nb$_{5}$, also Re$_{24}$Ti$_{5}$ adopts an $\alpha$-Mn type crystal structure with space group $I$-43$m$. However, while the former compounds have been widely  studied by means of macro- and microscopic techniques,\cite{Lue2011, Matano2016} much less is known about Re$_{24}$Ti$_{5}$. A simple analogy, based on structural similarity, can lead to wrong conclusions, since a SOC-dependent parity mixing can bring about rather different superconducting properties. Since its sister compounds, Re$_6$(Zr,Hf), are known to break the TRS in the superconducting state,\cite{Singh2014,Singh2017} Re$_{24}$Ti$_{5}$ represents an ideal opportunity to search for TRS breaking and unconventional SC in a material with a modified SOC value. Moreover, the study of an additional NCSCs can bring new insights into the nature of unconventional superconductivity in general.

Considering the key role played by muon-spin relaxation and rotation ($\mu$SR) techniques in unraveling the presence of TRS breaking in unconventional superconductors,\cite{Luke1998} in this paper, we report on the systematic magnetization, transport,
thermodynamic, tunnel-diode oscillator (TDO) and $\mu$SR studies of Re$_{24}$Ti$_{5}$, with 
particular focus on the latter. We find that below $T_c$
spontaneous magnetic fields appear, implying a superconducting state 
which breaks TRS and has an unconventional nature. The low-temperature penetration depth, superfluid density and electronic specific heat all suggest a nodeless $s$-wave pairing mechanism.
%

Polycrystalline Re$_{24}$Ti$_{5}$ samples were prepared by arc melting Re
and Ti metals under argon atmosphere and then annealed at 900$^\circ$C for two weeks. The x-ray
powder diffraction, measured on a Bruker D8 diffractometer,
confirmed the $\alpha$-Mn structure of Re$_{24}$Ti$_{5}$.
Magnetic susceptibility, electrical resistivity, and specific heat measurements in different applied magnetic
fields were performed on a Quantum Design magnetic property measurement
system (MPMS-7\,T) and a physical property measurement system (PPMS-14\,T).
The $\mu$SR measurements were carried out using the general-purpose
(GPS) instrument located at the $\pi$M3
beamline of the Swiss Muon Source (S$\mu$S) of Paul Scherrer
Institut (PSI) in Villigen, Switzerland. The temperature-dependent shift of
magnetic-penetration depth was measured by using a tunnel-diode 
oscillator (TDO) technique in a He$^3$ cryostat, at an operating frequency of 7\,MHz.
%
%

\begin{figure}[bth]
  \centering
  \includegraphics[width=0.48\textwidth]{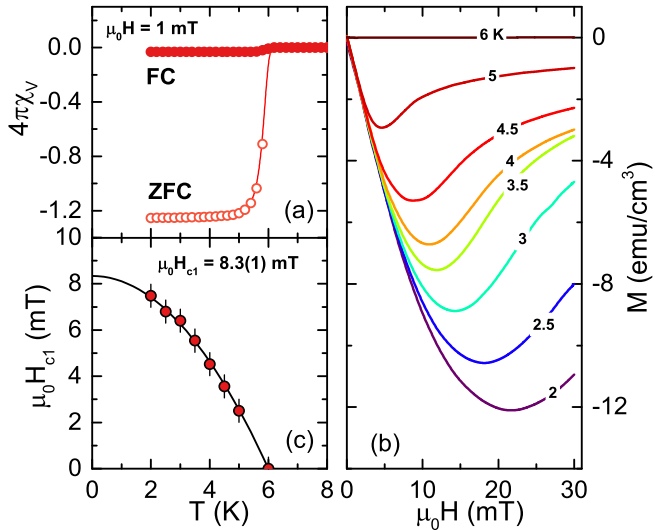}
  \caption{\label{fig:superconductivity} (a) Temperature dependence of magnetic susceptibility $\chi(T)$ for Re$_{24}$Ti$_{5}$. (b) Magnetization versus applied  magnetic field recorded at different temperatures up to $T_c$. For each temperature, $\mu_{0}H_{c1}$ was determined as the value where $M(H)$ deviates from linearity. (c) $\mu_{0}H_{c1}$ vs.\ temperature: the solid line, a fit to $\mu_{0}H_{c1}(T) =\mu_{0}H_{c1}(0)[1-(T/T_{c})^2]$, determines a $\mu_{0} H_{c1}(0) = 8.3(1)$\,mT.}
\end{figure}

\begin{figure}[htb]
  \centering
  \includegraphics[width=0.52\textwidth]{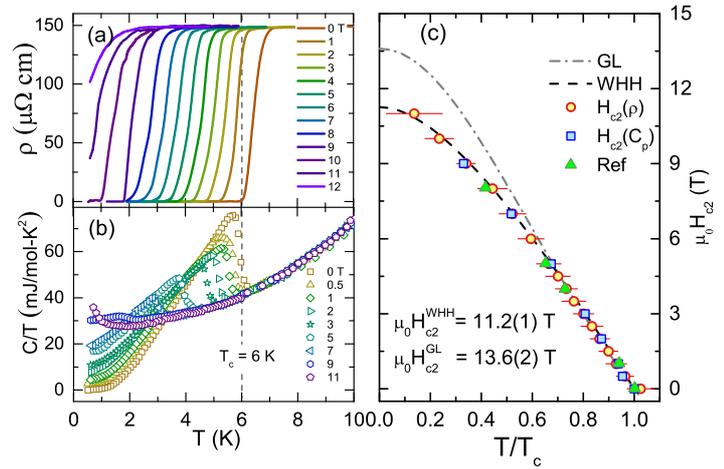}
  \caption{\label{fig:Hc2_determ} Temperature-dependent electrical resistivity (a)
  and specific heat (b) at different applied magnetic fields up to 12 T. From
  the suppression of $T_{c}$ with increasing field (c)
  we determine an upper critical field $\mu_{0}H_{c2}(0) = 11.2(1)$\,T.
  The dashed line represents a fit to the WHH model without spin-orbit
  scattering, whereas the dash-dotted line is a fit to the Ginzburg-Landau model (see text). }
\end{figure}

The magnetic susceptibility, measured at 1\,mT using field-cooling (FC) and zero-field cooling (ZFC) procedures,
is shown in Fig.~\ref{fig:superconductivity}(a). The splitting of the two curves is typical of type-II superconductors, and the
ZFC-susceptibility indicates bulk superconductivity with $T_c$ = 6\,K. The electrical resistivity drops at
the onset of superconductivity at 6.8\,K, becoming zero at 6\,K [see Fig.~\ref{fig:Hc2_determ}(a)]. The bulk
nature of SC is further confirmed by specific-heat data [see Fig.~\ref{fig:Hc2_determ}(b)].
%

\begin{figure}[th]
	\centering
	\includegraphics[width=0.4\textwidth,angle=0]{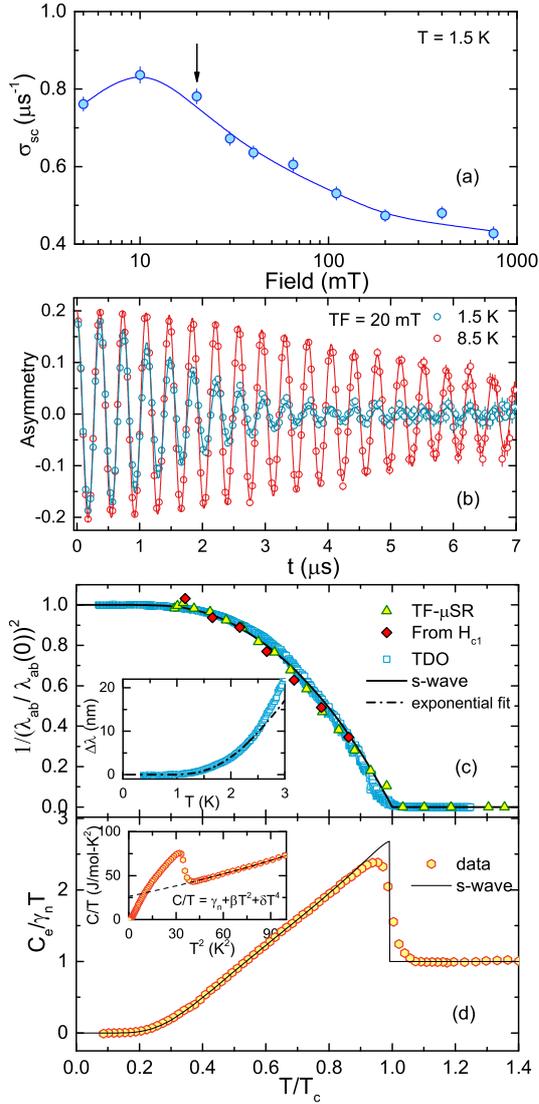}
	\vspace{-2ex}%
	\caption{\label{fig:TF_MuSR} (a) Field-dependent $\mu$SR relaxation rate at
		$T = 1.5$\,K. The arrow indicates the field used for the TF-$\mu$SR studies of
		the superconducting phase. (b) Time-domain TF-$\mu$SR spectra above
		and below $T_{c}$ show different relaxation rates. (c) Superfluid density vs.\ temperature, as determined from $\mu$SR
		(triangles), $H_{c1}$ (diamonds) and TDO (squares) data. The insert shows the shift of penetration depth below 3\,K and the dash-dotted line indicates the exponential temperature dependence. (d) Zero-field electronic specific heat vs.\ temperature. Inset: Raw $C/T$ data vs.\ $T^{2}$. The dashed-line is a fit to $C/T = \gamma_{\mathrm{n}} + \beta T^{2} + \delta T^{4}$, from which the phonon contribution was evaluated.
		The solid lines in (c) and (d) both represent fits using a fully-gapped $s$-wave model.}
\end{figure}

In transverse field (TF) $\mu$SR measurements of superconductors, 
the applied magnetic field should exceed the lower $\mu_{0}H_{c1}$ 
critical value, so that the
additional field-distribution broadening due to the flux-line lattice (FFL) can be quantified from the
muon decay rate. To determine  $\mu_{0}H_{c1}$, the field-dependent magnetization was 
preliminarily measured at various temperatures below $T_c$, as shown in
Fig.~\ref{fig:superconductivity}(b). The derived $\mu_{0}H_{c1}$ values are plotted in Fig.~\ref{fig:superconductivity}(c)
as a function of temperature. The solid line is a fit to
$\mu_{0}H_{c1}(T) = \mu_{0}H_{c1}(0)[1-(T/T_{c})^2]$,
which provides a lower critical field $\mu_{0} H_{c1}(0) = 8.3(1)$\,mT, consistent with the 8.4-mT value calculated from magnetic penetration depth $\lambda(0)$.
In the Ginzburg-Landau theory of superconductivity, the magnetic penetration
depth $\lambda$ is related to the coherence length $\xi$ and the lower
critical field $\mu_{0}H_{c1}$ via 
$\mu_{0}H_{c1} = (\Phi_0 /4 \pi \lambda^2)[$ln$(\kappa)+\alpha(\kappa)]$,
where $\Phi_0 = 2.07 \times 10^{-3}$\,T$\mu$m$^{2}$ is the quantum
of magnetic flux, $\kappa$ = $\lambda$/$\xi$ is the Ginzburg-Landau
parameter, and $\alpha(\kappa)$ is a 
parameter which converges to 0.497 for $\kappa$ $\gg$ 1. By using
$\mu_{0}H_{c1} = 8.3$\,mT and $\xi = 5.41$\,nm (calculated from $\mu_{0}H_{c2}$), the
resulting $\lambda(0) = 286$\,nm is consistent with the experimental
value from $\mu$SR [see Fig.~\ref{fig:TF_MuSR}(c)].
With a Ginzburg-Landau parameter $\kappa \sim 53 \gg 1$, Re$_{24}$Ti$_{5}$ is
clearly a type-II superconductor. The temperature dependence of penetration depth
$\lambda(T)$ can be estimated also from $\mu_{0}H_{c1}(T)$
and $\xi(T)$, where $\xi(T)$ is related to the upper critical field,
$\mu_{0} H_{c2}(T)$ = $\Phi_0$/2$\pi$$\xi^2(T)$.
%
%

To investigate the behavior of the upper critical field $\mu_0$$H_{c2}$,
we measured the electrical resistivity $\rho(T)$ and specific heat $C(T)$/$T$
at various magnetic fields. As shown in Figs.~\ref{fig:Hc2_determ}(a) and (b),
the superconducting transition in both cases shifts towards lower temperature upon increasing the magnetic field.
Note that, for $\mu_{0}H$ = 11\,T, the large upturn of specific heat at low-$T$ is due to a Schottky anomaly from nuclear moments, which
hides the superconducting transition. Similar features were also observed in other Re-based intermetallic superconductors.\cite{chen2016ReHf,chen2013ReNb}
The superconducting transition temperatures vs.\ the normalized temperature $T/T_{c}$, as
derived from both $\rho(T)$ and  $C(T)$/$T$ are summarized in Fig.~\ref{fig:Hc2_determ}(c).
Data taken from Ref.~\onlinecite{Lue2013ReTi} are also plotted. The temperature dependence
of the upper critical field $\mu_0$$H_{c2}(T)$ was analyzed
following the Werthamer-Helfand-Hohenberg (WHH) model.\cite{Werthamer1966}
The dashed-line in Fig.~\ref{fig:Hc2_determ}(c),  a fit to the WHH model
without considering spin-orbital scattering, gives $\mu_0 H_{c2}^\mathrm{WHH}(0) = \- 11.2(1)$\,T.
The derived $\mu_0$$H_{c2}(0)$ value is very close to the Pauli
paramagnetic limit for the weak-coupling case, 
$\mu_0 H_\mathrm{P} = 1.86\,T_{c} = 11.7(2)$\,T, thus indicating
the possibility of a singlet-triplet mixing state.
For completeness, we estimated the upper critical field also by means of
the Ginzburg-Landau model
$\mu_0 H_{c2}(T) = \mu_0 H_{c2}(0)(1-t^2)/(1+t^2)$,
where $t = T/T_{c}$ is again the normalized temperature.
As shown in Fig.~\ref{fig:Hc2_determ}(c) by a dash-dotted line,
at low fields the fit is quite good. However, at higher applied
fields, the fit deviates significantly from data, providing an
overestimated critical field value $\mu_0 H_{c2}^\mathrm{GL}(0) = 13.6(2)$\,T.
The remarkable agreement of the more elaborate WHH model with experimental data is clearly seen in Fig.~\ref{fig:Hc2_determ}(c).
%

To investigate the superconducting properties of Re$_{24}$Ti$_{5}$ at a microscopic level,
we carried out TF-$\mu$SR measurements in an applied field of 20\,mT.
The optimal field value for such experiments was determined
via a preliminary field-dependent $\mu$SR depolarization-rate measurement at 1.5\,K.
To avoid flux-pinning issues, the magnetic field
(up to 750\,mT) was applied in the normal state and then the sample was cooled down to 1.5\,K.
As shown in Fig.~\ref{fig:TF_MuSR}(a), the resulting Gaussian relaxation
rate $\sigma_\mathrm{sc}$ versus applied magnetic field exhibits a maximum
near the lower critical field [see Fig.~\ref{fig:superconductivity}(c)].
By considering the decrease of intervortex distance with field and
vortex-core effects, a field of 20\,mT(shown with an arrow), almost twice
the $\mu_{0} H_{c1}(0)$ value, was chosen for the temperature-dependent
study.

Figure~\ref{fig:TF_MuSR}(b) shows two representative TF-$\mu$SR
spectra collected above and below $T_{c}$. Below $T_{c}$,
the fast decay of muon-spin polarization reflects the inhomogeneous
field distribution due to the FFL in the mixed
superconducting state. The time-domain spectra were fitted by means of the following model with a Gaussian decay:
\begin{equation}
A_\mathrm{TF} = A_\mathrm{s} \cos(\gamma_{\mu} B_\mathrm{s} t + \phi) e^{- \sigma^2 t^2/2} +
A_\mathrm{bg} \cos(\gamma_{\mu} B_\mathrm{bg} t + \phi).
\end{equation}
Here $A_\mathrm{s}$ and $A_\mathrm{bg}$ are the initial muon-spin
asymmetries for muons implanted in the sample and sample holder,
respectively, with the latter not undergoing any depolarization.
$\gamma_{\mu} = 2\pi \times 135.53$\,MHz/T is the muon gyromagnetic ratio, $B_\mathrm{s}$ and $B_\mathrm{bg}$ are the local fields
sensed by implanted muons in the sample and sample holder,
$\phi$ is the (common) initial precession phase, and $\sigma$ is
a Gaussian-relaxation rate. Given the nonmagnetic nature of the sample
holder, $B_\mathrm{bg}$ practically coincides with the applied
magnetic field and was used as an intrinsic reference.

In the superconducting state, the Gaussian
relaxation rate includes contributions from both the FLL
($\sigma_\mathrm{sc}$) and the nuclear magnetic moments
($\sigma_\mathrm{n}$). Since $\sigma_\mathrm{n}$ is expected to
be temperature independent in the considered temperature range, the
FLL-related relaxation rate can be derived by subtracting the nuclear
contribution from the measured Gaussian relaxation, i.e., 
$\sigma_\mathrm{sc}$ = $\sqrt{\sigma^{2} - \sigma^{2}_\mathrm{n}}$.
Since $\sigma_\mathrm{sc}$ is directly related to the superfluid density
($\sigma_\mathrm{sc}$ $\propto$ $1/\lambda^2$), the superconducting gap
value and its symmetry can be determined from the temperature-dependent relaxation
rate $\sigma_\mathrm{sc}(T)$. For small applied magnetic fields [in comparison
with the upper critical field, i.e., $H_\mathrm{appl}$/$H_{c2}$ $\ll$\,1],
the effective penetration depth $\lambda_\mathrm{eff}$ can be calculated
from:\cite{Barford1988,Brandt2003}
\begin{equation}
\frac{\sigma_\mathrm{sc}^2(T)}{\gamma^2_{\mu}} = 0.00371\, \frac{\Phi_0^2}{\lambda^4_{\mathrm{eff}}(T)}.
\end{equation}
In a polycrystalline sample, the effective penetration depth
$\lambda_\mathrm{eff}$ is usually determined by the shortest penetration
depth $\lambda_{ab}$, the two being related via
$\lambda_\mathrm{eff} = 3^{1/4} \lambda_{ab}$.\cite{Fesenko1991}
Figure~\ref{fig:TF_MuSR}(c) shows the normalized superfluid density
($\rho_{\mathrm{sc}} \propto \lambda_{ab}^{-2}$) as a function
of temperature for Re$_{24}$Ti$_{5}$. The $\lambda_{ab}^{-2}$ data calculated from $\mu_{0} H_{c1}$ 
and those from TDO measurements are also plotted, both clearly consistent with the $\mu$SR results.
The temperature-dependent behavior of $\lambda_{ab}^{-2}$ is well described by an $s$-wave model with a single SC gap of about 1.08\,meV and a $\lambda(0)$ of 298\,nm. Such superconducting gap
is similar to that of other Re-based intermetallic superconductors, e.g., Re$_6$Zr
(1.21 meV),\cite{Singh2014, mayoh2017ReZr} Re$_6$Hf (0.94 meV),\cite{singh2016ReHf,chen2016ReHf} Re$_{24}$Nb$_5$ (0.89 meV).\cite{Lue2011} Also the
$2\Delta/\mathrm{k}_\mathrm{B}T_{c}$ values of these compounds [e.g.,
4.2(1) for Re$_{24}$Ti$_{5}$] are higher than 3.53, the value expected
for a weakly-coupled BCS superconductor, thus indicating moderately
strong electron-phonon couplings in these materials. Moreover, the low temperature penetration depth shown in the inset of Fig.~\ref{fig:TF_MuSR}(c), exhibits an exponential temperature dependence, providing further evidence of fully-gapped superconductivity in Re$_{24}$Ti$_5$.%

Since the specific heat in the superconducting state also offers insights into the
superconducting gap and its symmetry, the zero-field specific heat data were further analyzed. The electronic specific heat ($C_\mathrm{e}$/$T$) is obtained by subtracting the phonon contribution from the experimental data.
As shown in the inset of Fig.~\ref{fig:TF_MuSR}(d) by a dashed-line, the normal-state
specific heat is fitted with $C/T = \gamma_\mathrm{n} +\beta T^2 + \delta T^4$. 
The derived $C_\mathrm{e}/T$ is then divided by the normal-state electronic specific heat
coefficient, as shown in the main panel as a function of temperature. The solid line in Fig.~\ref{fig:TF_MuSR}(d) represents a fit  with $\gamma_\mathrm{n}$ = 26.4(2)\,mJ\,mol$^{-1}$K$^{-2}$ and a single isotropic
gap $\Delta(0)= 1.9(1)\,k_\mathrm{B}T_c$. It reproduces very well
the experimental data, while being consistent with the TF-$\mu$SR and TDO results [see Fig.~\ref{fig:TF_MuSR}(c)]. The ratio $\Delta C/\gamma$$T_{c}$ was found to be 1.4, 
consistent with previous data\cite{Lue2013ReTi} and in good agreement with the BCS-theory value of 1.43.
%
%

\begin{figure}[ht]
\centering
\includegraphics[width=0.50\textwidth,angle=0]{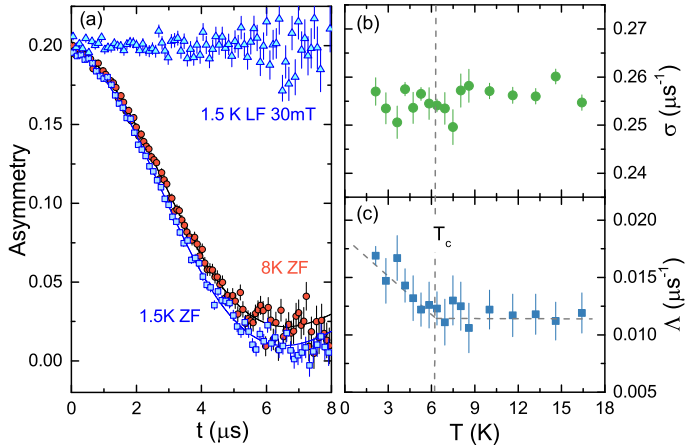}
\vspace{-2ex}%
\caption{\label{fig:ZF_muSR}(a) Representative zero-field $\mu$SR 
spectra for Re$_{24}$Ti$_{5}$ at 1.5\, and 8\,K and relevant 
fits by means of Eq.~(\ref{eq:KT_and_electr}). A typical
LF-$\mu$SR data set, collected at 1.5\,K in a 30-mT
longitudinal field, is also shown.  (b) Temperature dependence of the nuclear
relaxation rate $\sigma$, and (c) electronic relaxation rate $\Lambda$.
While $\sigma$ is almost temperature-independent, $\Lambda$ shows a
distinct increase below $T_c$.}
\end{figure}

To address the key question of the occurrence of time-reversal symmetry
breaking in Re$_{24}$Ti$_{5}$, we made use of zero-field (ZF)-$\mu$SR.
The large muon gyromagnetic ratio, combined with the availability
of 100\% spin-polarized muon beams, make ZF-$\mu$SR a very powerful
technique to detect the spontaneous fields, as shown by its
successful use in previous studies of Re$_{6}$(Zr,Hf),\cite{Singh2014,Singh2017}
La$_7$Ir$_3$,\cite{Barker2015}, Sr$_{2}$RuO$_{4}$,\cite{Luke1998}, and PrOs$_4$Sb$_{12}$.\cite{aoki2003}
Normally, in absence of external fields, the onset of superconducting
phase does not imply changes in the ZF muon-spin relaxation rate. However,
in case of TRS breaking, the onset of tiny spontaneous currents gives
rise to associated (weak) magnetic fields, promptly detected by ZF-$\mu$SR
as an increase in muon-spin relaxation rate.
Given the tiny size of such effects,  we measured carefully the muon-spin
relaxation rate both well above $T_c$  and well inside the  
superconducting phase. As shown in Fig.~\ref{fig:ZF_muSR}(a), two
representative ZF-$\mu$SR spectra collected above (8\,K) and below (1.5\,K) $T_c$
show clear differences, especially at long times. To exclude the possibility of stray
magnetic fields (which in any case would affect uniformly all datasets),
%
%
the magnets were quenched before the measurements and we
made use of an active field-nulling facility. Without an external field, the relaxation is determined mostly by the nuclear magnetic moments, normally described by a Gaussian Kubo-Toyabe relaxation function.\cite{Kubo1967,Yaouanc2011}
A possible spontaneous field contribution, is accounted for by an
additional exponential decay term. Consequently, the ZF-$\mu$SR spectra could be fitted by means of a combined Lorentzian and
Gaussian Kubo-Toyabe relaxation function:
\begin{equation}
\label{eq:KT_and_electr}
A_\mathrm{CKT} = A_\mathrm{s}\left[\frac{1}{3} + \frac{2}{3}(1 -
\sigma^{2}t^{2} - \Lambda t)\,
\mathrm{e}^{\left(-\frac{\sigma^{2}t^{2}}{2} - \Lambda t\right)} \right] + A_\mathrm{bg}.
\end{equation}
Here $A_\mathrm{s}$ is the initial sample-related muon-spin asymmetry,
whereas $A_\mathrm{bg}$ represents a time- and temperature-independent
background. As already shown in the TF-$\mu$SR case (see Fig.~\ref{fig:TF_MuSR}),
both the background and the nuclear contributions to the decay are independent of temperature.
This is clearly the case also with ZF-$\mu$SR [see
Fig.~\ref{fig:ZF_muSR}(b)], where $\sigma(T)$ remains constant (within the experimental error) in the studied temperature range.
On the other hand, the exponential component, related to the presence
of spontaneous magnetic fields, shows a small yet distinct increase
as the temperature is lowered below $T_c$ [see Fig.~\ref{fig:ZF_muSR}(c)].

Such an increase in $\Lambda(T)$, similar to that found also in 
Re$_{6}$(Zr,Hf),\cite{Singh2014,Singh2017} represents the signature of  
spontaneously occurring magnetic fields and of TRS breaking in 
the Re$_{24}$Ti$_{5}$ noncentrosymmetric superconductor.
Given the small size of the considered effect, to rule out the
possibility of an impurity-induced relaxation (typically relevant at low
temperatures), we performed auxiliary longitudinal-field
(LF)-$\mu$SR measurements at 1.5\,K. As
shown in Fig.~\ref{fig:ZF_muSR}(a), a 
field of 30\,mT only is sufficient to lock the muon spins
and to completely decouple them from the weak spontaneous
magnetic fields, thus removing any relaxation traces related to them.

Up to now, several NCSCs, including LaNiC$_2$,\cite{Hillier2009} Re$_6$\-(Zr,Hf),\cite{Singh2014,Singh2017} and La$_7$\-Ir$_3$\cite{Barker2015} have been found to exhibit a TRS breaking in the superconducting state. Yet, in many others, as e.g., Mo$_3$Al$_2$C,\cite{bauer2010} Mg$_{10}$Ir$_{19}$B$_{16}$,\cite{Acze2010} Re$_3$W,\cite{Biswas2012} and Pb\-Ta\-Se$_2$,\cite{Wilson2017} the TRS is preserved.
The Re$_{24}$Ti$_5$ considered here, a sister compound to Re$_6$(Zr,Hf), is a new member of the TRS-breaking NCSCs, 
despite a relatively reduced ASOC. This strongly suggests that, while the presence of an ASOC seems essential to induce a TRS breaking in NSSCs,  its strength is not a crucial condition. Indeed, although LaNiC$_2$\cite{Hillier2009}  has a much weaker ASOC 
compared to La$_7$Ir$_3$,\cite{Barker2015} the respective changes in zero-field muon relaxation rates are comparable ($\Delta$$\Lambda$ $\sim$ 0.01\,$\mu$s$^{-1}$). In our case, too, the replacement of the 5$d$ Hf with the 3$d$ Ti,  reduces remarkably the ASOC, yet the effects on TRS breaking remain comparable.  Hence, we believe that TRS breaking in NSCSs is mostly related to the crystal-structure symmetry and,  to test such hypothesis, La$_7$T$_3$ compounds (T = transition metal, e.g., Ni, Pd, Rh, Pt) represent good candidates, since all of them exhibit a Th$_7$Fe$_3$-type crystal structure, with the 3$d$ to 5$d$ transition metals covering a wide ASOC range.

The spin-triplet states can give rise to spontaneous fields in the superconducting state, which break the TRS. Most of these TRS-broken phases exhibit nodes in the superconducting gap, as e.g., Sr$_2$RuO$_4$.\cite{Luke1998} However, in highly-symmetric systems, the TRS breaking can also occur in fully-gapped states.\cite{Agterberg1999} Thus, the cubic Re$_6$(Zr,Hf)~\cite{Singh2014,Singh2017} and Re$_{24}$Ti$_{5}$ or the hexagonal La$_7$Ir$_3$~\cite{Barker2015} all exhibit fully gapped superconducting states, but with TRS breaking. A point-group analysis of Re$_6$Zr~\cite{Singh2014} reveals that a mixed singlet and triplet state is allowed to break the TRS. The continuous search for other low-symmetry NCSCs provides a good opportunity to find non-$s$-wave superconductors with TRS breaking, hence, furthering our understanding of the NCSC physics.

In summary, we investigated the noncentrosymmetric superconductor
Re$_{24}$Ti$_{5}$ by means of $\mu$SR and TDO techniques. Bulk superconductivity
with $T_c$ = 6\,K was characterized by magnetization, transport,
and specific heat measurements. Both the low-temperature penetration depth, superfluid
density and the zero-field specific-heat data reveal a
nodeless superconductivity in Re$_{24}$Ti$_{5}$, well described
by an isotropic $s$-wave model with a single gap.
The spontaneous fields, which appear below $T_c$ and
increase with decreasing temperature, provide strong evidence that
the superconducting state of noncentrosymmetric Re$_{24}$Ti$_{5}$ breaks TRS and has an unconventional nature.
%

This work was supported by the National Key R\&D Program of China (Grants No. 2017YFA0303100 and No. 2016YFA0300202), the National Natural Science Foundation of China (Grants No. 11474251), and  the
Schwei\-ze\-rische Na\-ti\-o\-nal\-fonds zur F\"{o}r\-de\-rung
der Wis\-sen\-schaft\-lich\-en For\-schung (SNF).

\bibliography{Re24Ti5_biblio}

\begin{thebibliography}{36}%
\makeatletter
\providecommand \@ifxundefined [1]{%
 \@ifx{#1\undefined}
}%
\providecommand \@ifnum [1]{%
 \ifnum #1\expandafter \@firstoftwo
 \else \expandafter \@secondoftwo
 \fi
}%
\providecommand \@ifx [1]{%
 \ifx #1\expandafter \@firstoftwo
 \else \expandafter \@secondoftwo
 \fi
}%
\providecommand \natexlab [1]{#1}%
\providecommand \enquote  [1]{``#1''}%
\providecommand \bibnamefont  [1]{#1}%
\providecommand \bibfnamefont [1]{#1}%
\providecommand \citenamefont [1]{#1}%
\providecommand \href@noop [0]{\@secondoftwo}%
\providecommand \href [0]{\begingroup \@sanitize@url \@href}%
\providecommand \@href[1]{\@@startlink{#1}\@@href}%
\providecommand \@@href[1]{\endgroup#1\@@endlink}%
\providecommand \@sanitize@url [0]{\catcode `\\12\catcode `\$12\catcode
  `\&12\catcode `\#12\catcode `\^12\catcode `\_12\catcode `\%12\relax}%
\providecommand \@@startlink[1]{}%
\providecommand \@@endlink[0]{}%
\providecommand \url  [0]{\begingroup\@sanitize@url \@url }%
\providecommand \@url [1]{\endgroup\@href {#1}{\urlprefix }}%
\providecommand \urlprefix  [0]{URL }%
\providecommand \Eprint [0]{\href }%
\providecommand \doibase [0]{http://dx.doi.org/}%
\providecommand \selectlanguage [0]{\@gobble}%
\providecommand \bibinfo  [0]{\@secondoftwo}%
\providecommand \bibfield  [0]{\@secondoftwo}%
\providecommand \translation [1]{[#1]}%
\providecommand \BibitemOpen [0]{}%
\providecommand \bibitemStop [0]{}%
\providecommand \bibitemNoStop [0]{.\EOS\space}%
\providecommand \EOS [0]{\spacefactor3000\relax}%
\providecommand \BibitemShut  [1]{\csname bibitem#1\endcsname}%
\let\auto@bib@innerbib\@empty
\bibitem [{\citenamefont {Bauer}\ and\ \citenamefont
  {Sigrist}(2012)}]{Bauer2012}%
  \BibitemOpen
  \bibinfo {editor} {\bibfnamefont {E.}~\bibnamefont {Bauer}}\ and\ \bibinfo
  {editor} {\bibfnamefont {M.}~\bibnamefont {Sigrist}},\ eds.,\ \href@noop {}
  {\emph {\bibinfo {title} {Non-Centrosymmetric Superconductors}}},\ Vol.\
  \bibinfo {volume} {847}\ (\bibinfo  {publisher} {Springer Verlag},\ \bibinfo
  {address} {Berlin},\ \bibinfo {year} {2012})\BibitemShut {NoStop}%
\bibitem [{\citenamefont {Smidman}\ \emph {et~al.}(2017)\citenamefont
  {Smidman}, \citenamefont {Salamon}, \citenamefont {Yuan},\ and\ \citenamefont
  {Agterberg}}]{smidman2017}%
  \BibitemOpen
  \bibfield  {author} {\bibinfo {author} {\bibfnamefont {M.}~\bibnamefont
  {Smidman}}, \bibinfo {author} {\bibfnamefont {M.~B.}\ \bibnamefont
  {Salamon}}, \bibinfo {author} {\bibfnamefont {H.~Q.}\ \bibnamefont {Yuan}}, \
  and\ \bibinfo {author} {\bibfnamefont {D.~F.}\ \bibnamefont {Agterberg}},\
  }\bibfield  {title} {\enquote {\bibinfo {title} {Superconductivity and
  spin--orbit coupling in non-centrosymmetric materials: {A} review},}\ }\href
  {http://stacks.iop.org/0034-4885/80/i=3/a=036501} {\bibfield  {journal}
  {\bibinfo  {journal} {Rep. Prog. Phys.}\ }\textbf {\bibinfo {volume} {80}},\
  \bibinfo {pages} {036501} (\bibinfo {year} {2017})}\BibitemShut {NoStop}%
\bibitem [{\citenamefont {Bonalde}\ \emph {et~al.}(2005)\citenamefont
  {Bonalde}, \citenamefont {Br{\"a}mer-Escamilla},\ and\ \citenamefont
  {Bauer}}]{bonalde2005CePt3Si}%
  \BibitemOpen
  \bibfield  {author} {\bibinfo {author} {\bibfnamefont {I.}~\bibnamefont
  {Bonalde}}, \bibinfo {author} {\bibfnamefont {W.}~\bibnamefont
  {Br{\"a}mer-Escamilla}}, \ and\ \bibinfo {author} {\bibfnamefont
  {E.}~\bibnamefont {Bauer}},\ }\bibfield  {title} {\enquote {\bibinfo {title}
  {Evidence for line nodes in the superconducting energy gap of
  noncentrosymmetric {Ce}{Pt}$_{3}${Si} from magnetic penetration depth
  measurements},}\ }\href {\doibase 10.1103/PhysRevLett.94.207002} {\bibfield
  {journal} {\bibinfo  {journal} {Phys. Rev. Lett.}\ }\textbf {\bibinfo
  {volume} {94}},\ \bibinfo {pages} {207002} (\bibinfo {year}
  {2005})}\BibitemShut {NoStop}%
\bibitem [{\citenamefont {Mukuda}\ \emph {et~al.}(2008)\citenamefont {Mukuda},
  \citenamefont {Fujii}, \citenamefont {Ohara}, \citenamefont {Harada},
  \citenamefont {Yashima}, \citenamefont {Kitaoka}, \citenamefont {Okuda},
  \citenamefont {Settai},\ and\ \citenamefont {Onuki}}]{mukuda2008CeIrSi3}%
  \BibitemOpen
  \bibfield  {author} {\bibinfo {author} {\bibfnamefont {H.}~\bibnamefont
  {Mukuda}}, \bibinfo {author} {\bibfnamefont {T.}~\bibnamefont {Fujii}},
  \bibinfo {author} {\bibfnamefont {T.}~\bibnamefont {Ohara}}, \bibinfo
  {author} {\bibfnamefont {A.}~\bibnamefont {Harada}}, \bibinfo {author}
  {\bibfnamefont {M.}~\bibnamefont {Yashima}}, \bibinfo {author} {\bibfnamefont
  {Y.}~\bibnamefont {Kitaoka}}, \bibinfo {author} {\bibfnamefont
  {Y.}~\bibnamefont {Okuda}}, \bibinfo {author} {\bibfnamefont
  {R.}~\bibnamefont {Settai}}, \ and\ \bibinfo {author} {\bibfnamefont
  {Y.}~\bibnamefont {Onuki}},\ }\bibfield  {title} {\enquote {\bibinfo {title}
  {Enhancement of superconducting transition temperature due to the strong
  antiferromagnetic spin fluctuations in the noncentrosymmetric heavy-fermion
  superconductor {Ce}{Ir}{Si}$_{3}$: A $^{29}${Si} {NMR} study under
  pressure},}\ }\href {\doibase 10.1103/PhysRevLett.100.107003} {\bibfield
  {journal} {\bibinfo  {journal} {Phys. Rev. Lett.}\ }\textbf {\bibinfo
  {volume} {100}},\ \bibinfo {pages} {107003} (\bibinfo {year}
  {2008})}\BibitemShut {NoStop}%
\bibitem [{\citenamefont {Yuan}\ \emph {et~al.}(2006)\citenamefont {Yuan},
  \citenamefont {Agterberg}, \citenamefont {Hayashi}, \citenamefont {Badica},
  \citenamefont {Vandervelde}, \citenamefont {Togano}, \citenamefont
  {Sigrist},\ and\ \citenamefont {Salamon}}]{yuan2006}%
  \BibitemOpen
  \bibfield  {author} {\bibinfo {author} {\bibfnamefont {H.~Q.}\ \bibnamefont
  {Yuan}}, \bibinfo {author} {\bibfnamefont {D.~F.}\ \bibnamefont {Agterberg}},
  \bibinfo {author} {\bibfnamefont {N.}~\bibnamefont {Hayashi}}, \bibinfo
  {author} {\bibfnamefont {P.}~\bibnamefont {Badica}}, \bibinfo {author}
  {\bibfnamefont {D.}~\bibnamefont {Vandervelde}}, \bibinfo {author}
  {\bibfnamefont {K.}~\bibnamefont {Togano}}, \bibinfo {author} {\bibfnamefont
  {M.}~\bibnamefont {Sigrist}}, \ and\ \bibinfo {author} {\bibfnamefont
  {M.~B.}\ \bibnamefont {Salamon}},\ }\bibfield  {title} {\enquote {\bibinfo
  {title} {S-wave spin-triplet order in superconductors without inversion
  symmetry: {Li}$_{2}${Pd}$_{3}${B} and {Li}$_{2}${Pt}$_{3}${B}},}\ }\href
  {\doibase 10.1103/PhysRevLett.97.017006} {\bibfield  {journal} {\bibinfo
  {journal} {Phys. Rev. Lett.}\ }\textbf {\bibinfo {volume} {97}},\ \bibinfo
  {pages} {017006} (\bibinfo {year} {2006})}\BibitemShut {NoStop}%
\bibitem [{\citenamefont {Nishiyama}\ \emph {et~al.}(2007)\citenamefont
  {Nishiyama}, \citenamefont {Inada},\ and\ \citenamefont
  {Zheng}}]{nishiyama2007}%
  \BibitemOpen
  \bibfield  {author} {\bibinfo {author} {\bibfnamefont {M.}~\bibnamefont
  {Nishiyama}}, \bibinfo {author} {\bibfnamefont {Y.}~\bibnamefont {Inada}}, \
  and\ \bibinfo {author} {\bibfnamefont {Guo-qing}\ \bibnamefont {Zheng}},\
  }\bibfield  {title} {\enquote {\bibinfo {title} {Spin triplet superconducting
  state due to broken inversion symmetry in {Li}$_{2}${Pt}$_{3}${B}},}\ }\href
  {\doibase 10.1103/PhysRevLett.98.047002} {\bibfield  {journal} {\bibinfo
  {journal} {Phys. Rev. Lett.}\ }\textbf {\bibinfo {volume} {98}},\ \bibinfo
  {pages} {047002} (\bibinfo {year} {2007})}\BibitemShut {NoStop}%
\bibitem [{\citenamefont {Bauer}\ \emph {et~al.}(2010)\citenamefont {Bauer},
  \citenamefont {Rogl}, \citenamefont {Chen}, \citenamefont {Khan},
  \citenamefont {Michor}, \citenamefont {Hilscher}, \citenamefont {Royanian},
  \citenamefont {Kumagai}, \citenamefont {Li}, \citenamefont {Li},
  \citenamefont {Podloucky},\ and\ \citenamefont {Rogl}}]{bauer2010}%
  \BibitemOpen
  \bibfield  {author} {\bibinfo {author} {\bibfnamefont {E.}~\bibnamefont
  {Bauer}}, \bibinfo {author} {\bibfnamefont {G.}~\bibnamefont {Rogl}},
  \bibinfo {author} {\bibfnamefont {Xing-Qiu}\ \bibnamefont {Chen}}, \bibinfo
  {author} {\bibfnamefont {R.~T.}\ \bibnamefont {Khan}}, \bibinfo {author}
  {\bibfnamefont {H.}~\bibnamefont {Michor}}, \bibinfo {author} {\bibfnamefont
  {G.}~\bibnamefont {Hilscher}}, \bibinfo {author} {\bibfnamefont
  {E.}~\bibnamefont {Royanian}}, \bibinfo {author} {\bibfnamefont
  {K.}~\bibnamefont {Kumagai}}, \bibinfo {author} {\bibfnamefont {D.~Z.}\
  \bibnamefont {Li}}, \bibinfo {author} {\bibfnamefont {Y.~Y.}\ \bibnamefont
  {Li}}, \bibinfo {author} {\bibfnamefont {R.}~\bibnamefont {Podloucky}}, \
  and\ \bibinfo {author} {\bibfnamefont {P.}~\bibnamefont {Rogl}},\ }\bibfield
  {title} {\enquote {\bibinfo {title} {Unconventional superconducting phase in
  the weakly correlated noncentrosymmetric {Mo}$_{3}${Al}$_{2}${C} compound},}\
  }\href {\doibase 10.1103/PhysRevB.82.064511} {\bibfield  {journal} {\bibinfo
  {journal} {Phys. Rev. B}\ }\textbf {\bibinfo {volume} {82}},\ \bibinfo
  {pages} {064511} (\bibinfo {year} {2010})}\BibitemShut {NoStop}%
\bibitem [{\citenamefont {Chen}\ \emph
  {et~al.}(2013{\natexlab{a}})\citenamefont {Chen}, \citenamefont {Jiao},
  \citenamefont {Zhang}, \citenamefont {Chen}, \citenamefont {Yang},
  \citenamefont {Nicklas}, \citenamefont {Steglich},\ and\ \citenamefont
  {Yuan}}]{chen2013}%
  \BibitemOpen
  \bibfield  {author} {\bibinfo {author} {\bibfnamefont {J.}~\bibnamefont
  {Chen}}, \bibinfo {author} {\bibfnamefont {L.}~\bibnamefont {Jiao}}, \bibinfo
  {author} {\bibfnamefont {J.~L.}\ \bibnamefont {Zhang}}, \bibinfo {author}
  {\bibfnamefont {Y.}~\bibnamefont {Chen}}, \bibinfo {author} {\bibfnamefont
  {L.}~\bibnamefont {Yang}}, \bibinfo {author} {\bibfnamefont {M.}~\bibnamefont
  {Nicklas}}, \bibinfo {author} {\bibfnamefont {F.}~\bibnamefont {Steglich}}, \
  and\ \bibinfo {author} {\bibfnamefont {H.~Q.}\ \bibnamefont {Yuan}},\
  }\bibfield  {title} {\enquote {\bibinfo {title} {Evidence for two-gap
  superconductivity in the non-centrosymmetric compound {La}{Ni}{C}$_{2}$},}\
  }\href {http://stacks.iop.org/1367-2630/15/i=5/a=053005} {\bibfield
  {journal} {\bibinfo  {journal} {New J. Phys.}\ }\textbf {\bibinfo {volume}
  {15}},\ \bibinfo {pages} {053005} (\bibinfo {year}
  {2013}{\natexlab{a}})}\BibitemShut {NoStop}%
\bibitem [{\citenamefont {Kuroiwa}\ \emph {et~al.}(2008)\citenamefont
  {Kuroiwa}, \citenamefont {Saura}, \citenamefont {Akimitsu}, \citenamefont
  {Hiraishi}, \citenamefont {Miyazaki}, \citenamefont {Satoh}, \citenamefont
  {Takeshita},\ and\ \citenamefont {Kadono}}]{kuroiwa2008}%
  \BibitemOpen
  \bibfield  {author} {\bibinfo {author} {\bibfnamefont {S.}~\bibnamefont
  {Kuroiwa}}, \bibinfo {author} {\bibfnamefont {Y.}~\bibnamefont {Saura}},
  \bibinfo {author} {\bibfnamefont {J.}~\bibnamefont {Akimitsu}}, \bibinfo
  {author} {\bibfnamefont {M.}~\bibnamefont {Hiraishi}}, \bibinfo {author}
  {\bibfnamefont {M.}~\bibnamefont {Miyazaki}}, \bibinfo {author}
  {\bibfnamefont {K.~H.}\ \bibnamefont {Satoh}}, \bibinfo {author}
  {\bibfnamefont {S.}~\bibnamefont {Takeshita}}, \ and\ \bibinfo {author}
  {\bibfnamefont {R.}~\bibnamefont {Kadono}},\ }\bibfield  {title} {\enquote
  {\bibinfo {title} {Multigap superconductivity in sesquicarbides
  {La}$_{2}${C}$_{3}$ and {Y}$_{2}${C}$_{3}$},}\ }\href {\doibase
  10.1103/PhysRevLett.100.097002} {\bibfield  {journal} {\bibinfo  {journal}
  {Phys. Rev. Lett.}\ }\textbf {\bibinfo {volume} {100}},\ \bibinfo {pages}
  {097002} (\bibinfo {year} {2008})}\BibitemShut {NoStop}%
\bibitem [{\citenamefont {Bauer}\ \emph {et~al.}(2004)\citenamefont {Bauer},
  \citenamefont {Hilscher}, \citenamefont {Michor}, \citenamefont {Paul},
  \citenamefont {Scheidt}, \citenamefont {Gribanov}, \citenamefont {Seropegin},
  \citenamefont {No{\"e}l}, \citenamefont {Sigrist},\ and\ \citenamefont
  {Rogl}}]{bauer2004}%
  \BibitemOpen
  \bibfield  {author} {\bibinfo {author} {\bibfnamefont {E.}~\bibnamefont
  {Bauer}}, \bibinfo {author} {\bibfnamefont {G.}~\bibnamefont {Hilscher}},
  \bibinfo {author} {\bibfnamefont {H.}~\bibnamefont {Michor}}, \bibinfo
  {author} {\bibfnamefont {C.}~\bibnamefont {Paul}}, \bibinfo {author}
  {\bibfnamefont {E.~W.}\ \bibnamefont {Scheidt}}, \bibinfo {author}
  {\bibfnamefont {A.}~\bibnamefont {Gribanov}}, \bibinfo {author}
  {\bibfnamefont {Y.}~\bibnamefont {Seropegin}}, \bibinfo {author}
  {\bibfnamefont {H.}~\bibnamefont {No{\"e}l}}, \bibinfo {author}
  {\bibfnamefont {M.}~\bibnamefont {Sigrist}}, \ and\ \bibinfo {author}
  {\bibfnamefont {P.}~\bibnamefont {Rogl}},\ }\bibfield  {title} {\enquote
  {\bibinfo {title} {Heavy fermion superconductivity and magnetic order in
  noncentrosymmetric {Ce}{Pt}$_{3}${Si}},}\ }\href {\doibase
  10.1103/PhysRevLett.92.027003} {\bibfield  {journal} {\bibinfo  {journal}
  {Phys. Rev. Lett.}\ }\textbf {\bibinfo {volume} {92}},\ \bibinfo {pages}
  {027003} (\bibinfo {year} {2004})}\BibitemShut {NoStop}%
\bibitem [{\citenamefont {Kimura}\ \emph {et~al.}(2007)\citenamefont {Kimura},
  \citenamefont {Ito}, \citenamefont {Aoki}, \citenamefont {Uji},\ and\
  \citenamefont {Terashima}}]{kimura2007}%
  \BibitemOpen
  \bibfield  {author} {\bibinfo {author} {\bibfnamefont {N.}~\bibnamefont
  {Kimura}}, \bibinfo {author} {\bibfnamefont {K.}~\bibnamefont {Ito}},
  \bibinfo {author} {\bibfnamefont {H.}~\bibnamefont {Aoki}}, \bibinfo {author}
  {\bibfnamefont {S.}~\bibnamefont {Uji}}, \ and\ \bibinfo {author}
  {\bibfnamefont {T.}~\bibnamefont {Terashima}},\ }\bibfield  {title} {\enquote
  {\bibinfo {title} {Extremely high upper critical magnetic field of the
  noncentrosymmetric heavy fermion superconductor {Ce}{Rh}{Si}$_{3}$},}\ }\href
  {\doibase 10.1103/PhysRevLett.98.197001} {\bibfield  {journal} {\bibinfo
  {journal} {Phys. Rev. Lett.}\ }\textbf {\bibinfo {volume} {98}},\ \bibinfo
  {pages} {197001} (\bibinfo {year} {2007})}\BibitemShut {NoStop}%
\bibitem [{\citenamefont {Sugitani}\ \emph {et~al.}(2006)\citenamefont
  {Sugitani}, \citenamefont {Okuda}, \citenamefont {Shishido}, \citenamefont
  {Yamada}, \citenamefont {Thamizhavel}, \citenamefont {Yamamoto},
  \citenamefont {Matsuda}, \citenamefont {Haga}, \citenamefont {Takeuchi},
  \citenamefont {Settai},\ and\ \citenamefont {Onuki}}]{sugitani2006}%
  \BibitemOpen
  \bibfield  {author} {\bibinfo {author} {\bibfnamefont {I.}~\bibnamefont
  {Sugitani}}, \bibinfo {author} {\bibfnamefont {Y.}~\bibnamefont {Okuda}},
  \bibinfo {author} {\bibfnamefont {H.}~\bibnamefont {Shishido}}, \bibinfo
  {author} {\bibfnamefont {T.}~\bibnamefont {Yamada}}, \bibinfo {author}
  {\bibfnamefont {A.}~\bibnamefont {Thamizhavel}}, \bibinfo {author}
  {\bibfnamefont {E.}~\bibnamefont {Yamamoto}}, \bibinfo {author}
  {\bibfnamefont {T.~D.}\ \bibnamefont {Matsuda}}, \bibinfo {author}
  {\bibfnamefont {Y.}~\bibnamefont {Haga}}, \bibinfo {author} {\bibfnamefont
  {T.}~\bibnamefont {Takeuchi}}, \bibinfo {author} {\bibfnamefont
  {R.}~\bibnamefont {Settai}}, \ and\ \bibinfo {author} {\bibfnamefont
  {Y.}~\bibnamefont {Onuki}},\ }\bibfield  {title} {\enquote {\bibinfo {title}
  {Pressure-induced heavy-fermion superconductivity in antiferromagnet
  {Ce}{Ir}{Si}$_{3}$ without inversion symmetry},}\ }\href
  {https://doi.org/10.1143/JPSJ.76.051009} {\bibfield  {journal} {\bibinfo
  {journal} {J. Phys. Soc. Jpn.}\ }\textbf {\bibinfo {volume} {75}},\ \bibinfo
  {pages} {043703} (\bibinfo {year} {2006})}\BibitemShut {NoStop}%
\bibitem [{\citenamefont {Hillier}\ \emph {et~al.}(2009)\citenamefont
  {Hillier}, \citenamefont {Quintanilla},\ and\ \citenamefont
  {Cywinski}}]{Hillier2009}%
  \BibitemOpen
  \bibfield  {author} {\bibinfo {author} {\bibfnamefont {A.~D.}\ \bibnamefont
  {Hillier}}, \bibinfo {author} {\bibfnamefont {J.}~\bibnamefont
  {Quintanilla}}, \ and\ \bibinfo {author} {\bibfnamefont {R.}~\bibnamefont
  {Cywinski}},\ }\bibfield  {title} {\enquote {\bibinfo {title} {Evidence for
  time-reversal symmetry breaking in the noncentrosymmetric superconductor
  {LaNiC$_{2}$}},}\ }\href {\doibase 10.1103/PhysRevLett.102.117007} {\bibfield
   {journal} {\bibinfo  {journal} {Phys. Rev. Lett.}\ }\textbf {\bibinfo
  {volume} {102}},\ \bibinfo {pages} {117007} (\bibinfo {year}
  {2009})}\BibitemShut {NoStop}%
\bibitem [{\citenamefont {Singh}\ \emph {et~al.}(2014)\citenamefont {Singh},
  \citenamefont {Hillier}, \citenamefont {Mazidian}, \citenamefont
  {Quintanilla}, \citenamefont {Annett}, \citenamefont {Paul}, \citenamefont
  {Balakrishnan},\ and\ \citenamefont {Lees}}]{Singh2014}%
  \BibitemOpen
  \bibfield  {author} {\bibinfo {author} {\bibfnamefont {R.~P.}\ \bibnamefont
  {Singh}}, \bibinfo {author} {\bibfnamefont {A.~D.}\ \bibnamefont {Hillier}},
  \bibinfo {author} {\bibfnamefont {B.}~\bibnamefont {Mazidian}}, \bibinfo
  {author} {\bibfnamefont {J.}~\bibnamefont {Quintanilla}}, \bibinfo {author}
  {\bibfnamefont {J.~F.}\ \bibnamefont {Annett}}, \bibinfo {author}
  {\bibfnamefont {D.~McK.}\ \bibnamefont {Paul}}, \bibinfo {author}
  {\bibfnamefont {G.}~\bibnamefont {Balakrishnan}}, \ and\ \bibinfo {author}
  {\bibfnamefont {M.~R.}\ \bibnamefont {Lees}},\ }\bibfield  {title} {\enquote
  {\bibinfo {title} {Detection of time-reversal symmetry breaking in the
  noncentrosymmetric superconductor {Re}$_{6}${Zr} using muon-spin
  spectroscopy},}\ }\href {\doibase 10.1103/PhysRevLett.112.107002} {\bibfield
  {journal} {\bibinfo  {journal} {Phys. Rev. Lett.}\ }\textbf {\bibinfo
  {volume} {112}},\ \bibinfo {pages} {107002} (\bibinfo {year}
  {2014})}\BibitemShut {NoStop}%
\bibitem [{\citenamefont {Singh}\ \emph
  {et~al.}(2017{\natexlab{a}})\citenamefont {Singh}, \citenamefont {Barker},
  \citenamefont {Thamizhavel}, \citenamefont {Paul}, \citenamefont {Hillier},\
  and\ \citenamefont {Singh}}]{Singh2017}%
  \BibitemOpen
  \bibfield  {author} {\bibinfo {author} {\bibfnamefont {D.}~\bibnamefont
  {Singh}}, \bibinfo {author} {\bibfnamefont {J.~A.~T.}\ \bibnamefont
  {Barker}}, \bibinfo {author} {\bibfnamefont {A.}~\bibnamefont {Thamizhavel}},
  \bibinfo {author} {\bibfnamefont {D.~McK.}\ \bibnamefont {Paul}}, \bibinfo
  {author} {\bibfnamefont {A.~D.}\ \bibnamefont {Hillier}}, \ and\ \bibinfo
  {author} {\bibfnamefont {R.~P.}\ \bibnamefont {Singh}},\ }\bibfield  {title}
  {\enquote {\bibinfo {title} {Time-reversal symmetry breaking in the
  noncentrosymmetric superconductor {Re}$_{6}${Hf}: Further evidence for
  unconventional behavior in the $\alpha$-{Mn} family of materials},}\ }\href
  {\doibase 10.1103/PhysRevB.96.180501} {\bibfield  {journal} {\bibinfo
  {journal} {Phys. Rev. B}\ }\textbf {\bibinfo {volume} {96}},\ \bibinfo
  {pages} {180501} (\bibinfo {year} {2017}{\natexlab{a}})}\BibitemShut
  {NoStop}%
\bibitem [{\citenamefont {Barker}\ \emph {et~al.}(2015)\citenamefont {Barker},
  \citenamefont {Singh}, \citenamefont {Thamizhavel}, \citenamefont {Hillier},
  \citenamefont {Lees}, \citenamefont {Balakrishnan}, \citenamefont {Paul},\
  and\ \citenamefont {Singh}}]{Barker2015}%
  \BibitemOpen
  \bibfield  {author} {\bibinfo {author} {\bibfnamefont {J.~A.~T.}\
  \bibnamefont {Barker}}, \bibinfo {author} {\bibfnamefont {D.}~\bibnamefont
  {Singh}}, \bibinfo {author} {\bibfnamefont {A.}~\bibnamefont {Thamizhavel}},
  \bibinfo {author} {\bibfnamefont {A.~D.}\ \bibnamefont {Hillier}}, \bibinfo
  {author} {\bibfnamefont {M.~R.}\ \bibnamefont {Lees}}, \bibinfo {author}
  {\bibfnamefont {G.}~\bibnamefont {Balakrishnan}}, \bibinfo {author}
  {\bibfnamefont {D.~McK.}\ \bibnamefont {Paul}}, \ and\ \bibinfo {author}
  {\bibfnamefont {R.~P.}\ \bibnamefont {Singh}},\ }\bibfield  {title} {\enquote
  {\bibinfo {title} {Unconventional superconductivity in {La}$_{7}${Ir}$_{3}$
  revealed by muon spin relaxation: Introducing a new family of
  noncentrosymmetric superconductor that breaks time-reversal symmetry},}\
  }\href {\doibase 10.1103/PhysRevLett.115.267001} {\bibfield  {journal}
  {\bibinfo  {journal} {Phys. Rev. Lett.}\ }\textbf {\bibinfo {volume} {115}},\
  \bibinfo {pages} {267001} (\bibinfo {year} {2015})}\BibitemShut {NoStop}%
\bibitem [{\citenamefont {Matthias}\ \emph {et~al.}(1961)\citenamefont
  {Matthias}, \citenamefont {Compton},\ and\ \citenamefont
  {Corenzwit}}]{Matthias1961}%
  \BibitemOpen
  \bibfield  {author} {\bibinfo {author} {\bibfnamefont {B.~T.}\ \bibnamefont
  {Matthias}}, \bibinfo {author} {\bibfnamefont {V.~B.}\ \bibnamefont
  {Compton}}, \ and\ \bibinfo {author} {\bibfnamefont {E.}~\bibnamefont
  {Corenzwit}},\ }\bibfield  {title} {\enquote {\bibinfo {title} {Some new
  superconducting compounds},}\ }\href {\doibase
  http://dx.doi.org/10.1016/0022-3697(61)90066-X} {\bibfield  {journal}
  {\bibinfo  {journal} {J. Phys. Chem. Solids}\ }\textbf {\bibinfo {volume}
  {19}},\ \bibinfo {pages} {130} (\bibinfo {year} {1961})}\BibitemShut
  {NoStop}%
\bibitem [{\citenamefont {Lue}\ \emph {et~al.}(2013)\citenamefont {Lue},
  \citenamefont {Liu}, \citenamefont {Kuo}, \citenamefont {Shih}, \citenamefont
  {Lin}, \citenamefont {Kuo}, \citenamefont {Chu}, \citenamefont {Hung},\ and\
  \citenamefont {Chen}}]{Lue2013ReTi}%
  \BibitemOpen
  \bibfield  {author} {\bibinfo {author} {\bibfnamefont {C.~S.}\ \bibnamefont
  {Lue}}, \bibinfo {author} {\bibfnamefont {H.~F.}\ \bibnamefont {Liu}},
  \bibinfo {author} {\bibfnamefont {C.~N.}\ \bibnamefont {Kuo}}, \bibinfo
  {author} {\bibfnamefont {P.~S.}\ \bibnamefont {Shih}}, \bibinfo {author}
  {\bibfnamefont {J.~Y.}\ \bibnamefont {Lin}}, \bibinfo {author} {\bibfnamefont
  {Y.~K.}\ \bibnamefont {Kuo}}, \bibinfo {author} {\bibfnamefont {M.~W.}\
  \bibnamefont {Chu}}, \bibinfo {author} {\bibfnamefont {T.~L.}\ \bibnamefont
  {Hung}}, \ and\ \bibinfo {author} {\bibfnamefont {Y.~Y.}\ \bibnamefont
  {Chen}},\ }\bibfield  {title} {\enquote {\bibinfo {title} {Investigation of
  normal and superconducting states in noncentrosymmetric
  {Re}$_{24}${Ti}$_{5}$},}\ }\href
  {http://stacks.iop.org/0953-2048/26/i=5/a=055011} {\bibfield  {journal}
  {\bibinfo  {journal} {Supercond. Sci. Tech.}\ }\textbf {\bibinfo {volume}
  {26}},\ \bibinfo {pages} {055011} (\bibinfo {year} {2013})}\BibitemShut
  {NoStop}%
\bibitem [{\citenamefont {Lue}\ \emph {et~al.}(2011)\citenamefont {Lue},
  \citenamefont {Su}, \citenamefont {Liu},\ and\ \citenamefont
  {Young}}]{Lue2011}%
  \BibitemOpen
  \bibfield  {author} {\bibinfo {author} {\bibfnamefont {C.~S.}\ \bibnamefont
  {Lue}}, \bibinfo {author} {\bibfnamefont {T.~H.}\ \bibnamefont {Su}},
  \bibinfo {author} {\bibfnamefont {H.~F.}\ \bibnamefont {Liu}}, \ and\
  \bibinfo {author} {\bibfnamefont {Ben-Li}\ \bibnamefont {Young}},\ }\bibfield
   {title} {\enquote {\bibinfo {title} {Evidence for $s$-wave superconductivity
  in noncentrosymmetric {Re}$_{24}${Nb}$_{5}$ from ${}^{93}${Nb} {NMR}
  measurements},}\ }\href {\doibase 10.1103/PhysRevB.84.052509} {\bibfield
  {journal} {\bibinfo  {journal} {Phys. Rev. B}\ }\textbf {\bibinfo {volume}
  {84}},\ \bibinfo {pages} {052509} (\bibinfo {year} {2011})}\BibitemShut
  {NoStop}%
\bibitem [{\citenamefont {Matano}\ \emph {et~al.}(2016)\citenamefont {Matano},
  \citenamefont {Yatagai}, \citenamefont {Maeda},\ and\ \citenamefont
  {Zheng}}]{Matano2016}%
  \BibitemOpen
  \bibfield  {author} {\bibinfo {author} {\bibfnamefont {K.}~\bibnamefont
  {Matano}}, \bibinfo {author} {\bibfnamefont {R.}~\bibnamefont {Yatagai}},
  \bibinfo {author} {\bibfnamefont {S.}~\bibnamefont {Maeda}}, \ and\ \bibinfo
  {author} {\bibfnamefont {Guo-qing}\ \bibnamefont {Zheng}},\ }\bibfield
  {title} {\enquote {\bibinfo {title} {Full-gap superconductivity in
  noncentrosymmetric {Re}$_{6}${Zr}, {Re}$_{27}${Zr}$_{5}$, and
  {Re}$_{24}${Zr}$_{5}$},}\ }\href {\doibase 10.1103/PhysRevB.94.214513}
  {\bibfield  {journal} {\bibinfo  {journal} {Phys. Rev. B}\ }\textbf {\bibinfo
  {volume} {94}},\ \bibinfo {pages} {214513} (\bibinfo {year}
  {2016})}\BibitemShut {NoStop}%
\bibitem [{\citenamefont {Luke}\ \emph {et~al.}(1998)\citenamefont {Luke},
  \citenamefont {Fudamoto}, \citenamefont {Kojima}, \citenamefont {Larkin},
  \citenamefont {Merrin}, \citenamefont {Nachumi}, \citenamefont {Uemura},
  \citenamefont {Maeno}, \citenamefont {Mao}, \citenamefont {Mori},
  \citenamefont {Nakamura},\ and\ \citenamefont {Sigrist}}]{Luke1998}%
  \BibitemOpen
  \bibfield  {author} {\bibinfo {author} {\bibfnamefont {G.~M.}\ \bibnamefont
  {Luke}}, \bibinfo {author} {\bibfnamefont {Y.}~\bibnamefont {Fudamoto}},
  \bibinfo {author} {\bibfnamefont {K.~M.}\ \bibnamefont {Kojima}}, \bibinfo
  {author} {\bibfnamefont {M.~I.}\ \bibnamefont {Larkin}}, \bibinfo {author}
  {\bibfnamefont {J.}~\bibnamefont {Merrin}}, \bibinfo {author} {\bibfnamefont
  {B.}~\bibnamefont {Nachumi}}, \bibinfo {author} {\bibfnamefont {Y.~J.}\
  \bibnamefont {Uemura}}, \bibinfo {author} {\bibfnamefont {Y.}~\bibnamefont
  {Maeno}}, \bibinfo {author} {\bibfnamefont {Z.~Q.}\ \bibnamefont {Mao}},
  \bibinfo {author} {\bibfnamefont {Y.}~\bibnamefont {Mori}}, \bibinfo {author}
  {\bibfnamefont {H.}~\bibnamefont {Nakamura}}, \ and\ \bibinfo {author}
  {\bibfnamefont {M.}~\bibnamefont {Sigrist}},\ }\bibfield  {title} {\enquote
  {\bibinfo {title} {Time-reversal symmetry-breaking superconductivity in
  {Sr}$_{2}${RuO}$_{4}$},}\ }\href {\doibase 10.1038/29038} {\bibfield
  {journal} {\bibinfo  {journal} {Nature}\ }\textbf {\bibinfo {volume} {394}},\
  \bibinfo {pages} {558} (\bibinfo {year} {1998})}\BibitemShut {NoStop}%
\bibitem [{\citenamefont {Chen}\ \emph {et~al.}(2016)\citenamefont {Chen},
  \citenamefont {Guo}, \citenamefont {Wang}, \citenamefont {Su}, \citenamefont
  {Mao}, \citenamefont {Du}, \citenamefont {Zhou}, \citenamefont {Yang},\ and\
  \citenamefont {Fang}}]{chen2016ReHf}%
  \BibitemOpen
  \bibfield  {author} {\bibinfo {author} {\bibfnamefont {B.}~\bibnamefont
  {Chen}}, \bibinfo {author} {\bibfnamefont {Y.}~\bibnamefont {Guo}}, \bibinfo
  {author} {\bibfnamefont {H.}~\bibnamefont {Wang}}, \bibinfo {author}
  {\bibfnamefont {Q.}~\bibnamefont {Su}}, \bibinfo {author} {\bibfnamefont
  {Q.}~\bibnamefont {Mao}}, \bibinfo {author} {\bibfnamefont {J.}~\bibnamefont
  {Du}}, \bibinfo {author} {\bibfnamefont {Y.}~\bibnamefont {Zhou}}, \bibinfo
  {author} {\bibfnamefont {J.}~\bibnamefont {Yang}}, \ and\ \bibinfo {author}
  {\bibfnamefont {M.}~\bibnamefont {Fang}},\ }\bibfield  {title} {\enquote
  {\bibinfo {title} {Superconductivity in the noncentrosymmetric compound
  {Re}$_{6}${Hf}},}\ }\href {\doibase 10.1103/PhysRevB.94.024518} {\bibfield
  {journal} {\bibinfo  {journal} {Phys. Rev. B}\ }\textbf {\bibinfo {volume}
  {94}},\ \bibinfo {pages} {024518} (\bibinfo {year} {2016})}\BibitemShut
  {NoStop}%
\bibitem [{\citenamefont {Chen}\ \emph
  {et~al.}(2013{\natexlab{b}})\citenamefont {Chen}, \citenamefont {Jiao},
  \citenamefont {Zhang}, \citenamefont {Chen}, \citenamefont {Yang},
  \citenamefont {Nicklas}, \citenamefont {Steglich},\ and\ \citenamefont
  {Yuan}}]{chen2013ReNb}%
  \BibitemOpen
  \bibfield  {author} {\bibinfo {author} {\bibfnamefont {J.}~\bibnamefont
  {Chen}}, \bibinfo {author} {\bibfnamefont {L.}~\bibnamefont {Jiao}}, \bibinfo
  {author} {\bibfnamefont {J.~L.}\ \bibnamefont {Zhang}}, \bibinfo {author}
  {\bibfnamefont {Y.}~\bibnamefont {Chen}}, \bibinfo {author} {\bibfnamefont
  {L.}~\bibnamefont {Yang}}, \bibinfo {author} {\bibfnamefont {M.}~\bibnamefont
  {Nicklas}}, \bibinfo {author} {\bibfnamefont {F.}~\bibnamefont {Steglich}}, \
  and\ \bibinfo {author} {\bibfnamefont {H.~Q.}\ \bibnamefont {Yuan}},\
  }\bibfield  {title} {\enquote {\bibinfo {title} {{BCS}-like superconductivity
  in the noncentrosymmetric compounds {Nb}$_{x}${Re}$_{1-x}$ (${0.13}\leqslant
  {x} \leqslant {0.38}$)},}\ }\href {\doibase 10.1103/PhysRevB.88.144510}
  {\bibfield  {journal} {\bibinfo  {journal} {Phys. Rev. B}\ }\textbf {\bibinfo
  {volume} {88}},\ \bibinfo {pages} {144510} (\bibinfo {year}
  {2013}{\natexlab{b}})}\BibitemShut {NoStop}%
\bibitem [{\citenamefont {Werthamer}\ \emph {et~al.}(1966)\citenamefont
  {Werthamer}, \citenamefont {Helfand},\ and\ \citenamefont
  {Hohenberg}}]{Werthamer1966}%
  \BibitemOpen
  \bibfield  {author} {\bibinfo {author} {\bibfnamefont {N.~R.}\ \bibnamefont
  {Werthamer}}, \bibinfo {author} {\bibfnamefont {E.}~\bibnamefont {Helfand}},
  \ and\ \bibinfo {author} {\bibfnamefont {P.~C.}\ \bibnamefont {Hohenberg}},\
  }\bibfield  {title} {\enquote {\bibinfo {title} {Temperature and purity
  dependence of the superconducting critical field, ${{H}}_{c2}$. {III}.
  {E}lectron spin and spin-orbit effects},}\ }\href {\doibase
  10.1103/PhysRev.147.295} {\bibfield  {journal} {\bibinfo  {journal} {Phys.
  Rev.}\ }\textbf {\bibinfo {volume} {147}},\ \bibinfo {pages} {295} (\bibinfo
  {year} {1966})}\BibitemShut {NoStop}%
\bibitem [{\citenamefont {Barford}\ and\ \citenamefont
  {Gunn}(1988)}]{Barford1988}%
  \BibitemOpen
  \bibfield  {author} {\bibinfo {author} {\bibfnamefont {W.}~\bibnamefont
  {Barford}}\ and\ \bibinfo {author} {\bibfnamefont {J.~M.~F.}\ \bibnamefont
  {Gunn}},\ }\bibfield  {title} {\enquote {\bibinfo {title} {The theory of the
  measurement of the {L}ondon penetration depth in uniaxial type {II}
  superconductors by muon spin rotation},}\ }\href {\doibase
  http://dx.doi.org/10.1016/0921-4534(88)90014-7} {\bibfield  {journal}
  {\bibinfo  {journal} {Physica C}\ }\textbf {\bibinfo {volume} {156}},\
  \bibinfo {pages} {515} (\bibinfo {year} {1988})}\BibitemShut {NoStop}%
\bibitem [{\citenamefont {Brandt}(2003)}]{Brandt2003}%
  \BibitemOpen
  \bibfield  {author} {\bibinfo {author} {\bibfnamefont {E.~H.}\ \bibnamefont
  {Brandt}},\ }\bibfield  {title} {\enquote {\bibinfo {title} {Properties of
  the ideal {G}inzburg-{L}andau vortex lattice},}\ }\href {\doibase
  10.1103/PhysRevB.68.054506} {\bibfield  {journal} {\bibinfo  {journal} {Phys.
  Rev. B}\ }\textbf {\bibinfo {volume} {68}},\ \bibinfo {pages} {054506}
  (\bibinfo {year} {2003})}\BibitemShut {NoStop}%
\bibitem [{\citenamefont {Fesenko}\ \emph {et~al.}(1991)\citenamefont
  {Fesenko}, \citenamefont {Gorbunov},\ and\ \citenamefont
  {Smilga}}]{Fesenko1991}%
  \BibitemOpen
  \bibfield  {author} {\bibinfo {author} {\bibfnamefont {V.~I.}\ \bibnamefont
  {Fesenko}}, \bibinfo {author} {\bibfnamefont {V.~N.}\ \bibnamefont
  {Gorbunov}}, \ and\ \bibinfo {author} {\bibfnamefont {V.~P.}\ \bibnamefont
  {Smilga}},\ }\bibfield  {title} {\enquote {\bibinfo {title} {Analytical
  properties of muon polarization spectra in type-{II} superconductors and
  experimental data interpretation for mono- and polycrystalline {HTSCs}},}\
  }\href {\doibase http://dx.doi.org/10.1016/0921-4534(91)90063-5} {\bibfield
  {journal} {\bibinfo  {journal} {Physica C}\ }\textbf {\bibinfo {volume}
  {176}},\ \bibinfo {pages} {551--558} (\bibinfo {year} {1991})}\BibitemShut
  {NoStop}%
\bibitem [{\citenamefont {Mayoh}\ \emph {et~al.}(2017)\citenamefont {Mayoh},
  \citenamefont {Barker}, \citenamefont {Singh}, \citenamefont {Balakrishnan},
  \citenamefont {Paul},\ and\ \citenamefont {Lees}}]{mayoh2017ReZr}%
  \BibitemOpen
  \bibfield  {author} {\bibinfo {author} {\bibfnamefont {D.~A.}\ \bibnamefont
  {Mayoh}}, \bibinfo {author} {\bibfnamefont {J.~A.~T.}\ \bibnamefont
  {Barker}}, \bibinfo {author} {\bibfnamefont {R.~P.}\ \bibnamefont {Singh}},
  \bibinfo {author} {\bibfnamefont {G.}~\bibnamefont {Balakrishnan}}, \bibinfo
  {author} {\bibfnamefont {D.~Mck.}\ \bibnamefont {Paul}}, \ and\ \bibinfo
  {author} {\bibfnamefont {M.~R.}\ \bibnamefont {Lees}},\ }\bibfield  {title}
  {\enquote {\bibinfo {title} {Superconducting and normal-state properties of
  the noncentrosymmetric superconductor {Re}$_{6}${Zr}},}\ }\href {\doibase
  10.1103/PhysRevB.96.064521} {\bibfield  {journal} {\bibinfo  {journal} {Phys.
  Rev. B}\ }\textbf {\bibinfo {volume} {96}},\ \bibinfo {pages} {064521}
  (\bibinfo {year} {2017})}\BibitemShut {NoStop}%
\bibitem [{\citenamefont {Singh}\ \emph
  {et~al.}(2017{\natexlab{b}})\citenamefont {Singh}, \citenamefont {Hillier},
  \citenamefont {Thamizhavel},\ and\ \citenamefont {Singh}}]{singh2016ReHf}%
  \BibitemOpen
  \bibfield  {author} {\bibinfo {author} {\bibfnamefont {D.}~\bibnamefont
  {Singh}}, \bibinfo {author} {\bibfnamefont {A.~D.}\ \bibnamefont {Hillier}},
  \bibinfo {author} {\bibfnamefont {A.}~\bibnamefont {Thamizhavel}}, \ and\
  \bibinfo {author} {\bibfnamefont {R.~P.}\ \bibnamefont {Singh}},\ }\bibfield
  {title} {\enquote {\bibinfo {title} {Superconducting properties of the
  noncentrosymmetric superconductor {Re}$_{6}${Hf}},}\ }\href {\doibase
  10.1103/PhysRevB.96.064521} {\bibfield  {journal} {\bibinfo  {journal} {Phys.
  Rev. B}\ }\textbf {\bibinfo {volume} {96}},\ \bibinfo {pages} {064521}
  (\bibinfo {year} {2017}{\natexlab{b}})}\BibitemShut {NoStop}%
\bibitem [{\citenamefont {Aoki}\ \emph {et~al.}(2003)\citenamefont {Aoki},
  \citenamefont {Tsuchiya}, \citenamefont {Kanayama}, \citenamefont {Saha},
  \citenamefont {Sugawara}, \citenamefont {Sato}, \citenamefont {Higemoto},
  \citenamefont {Koda}, \citenamefont {Ohishi}, \citenamefont {Nishiyama},\
  and\ \citenamefont {Kadono}}]{aoki2003}%
  \BibitemOpen
  \bibfield  {author} {\bibinfo {author} {\bibfnamefont {Y.}~\bibnamefont
  {Aoki}}, \bibinfo {author} {\bibfnamefont {A.}~\bibnamefont {Tsuchiya}},
  \bibinfo {author} {\bibfnamefont {T.}~\bibnamefont {Kanayama}}, \bibinfo
  {author} {\bibfnamefont {S.~R.}\ \bibnamefont {Saha}}, \bibinfo {author}
  {\bibfnamefont {H.}~\bibnamefont {Sugawara}}, \bibinfo {author}
  {\bibfnamefont {H.}~\bibnamefont {Sato}}, \bibinfo {author} {\bibfnamefont
  {W.}~\bibnamefont {Higemoto}}, \bibinfo {author} {\bibfnamefont
  {A.}~\bibnamefont {Koda}}, \bibinfo {author} {\bibfnamefont {K.}~\bibnamefont
  {Ohishi}}, \bibinfo {author} {\bibfnamefont {K.}~\bibnamefont {Nishiyama}}, \
  and\ \bibinfo {author} {\bibfnamefont {R.}~\bibnamefont {Kadono}},\
  }\bibfield  {title} {\enquote {\bibinfo {title} {Time-reversal
  symmetry-breaking superconductivity in heavy-fermion
  {Pr}{Os}$_{4}${Sb}$_{12}$ detected by muon-spin relaxation},}\ }\href
  {\doibase 10.1103/PhysRevLett.91.067003} {\bibfield  {journal} {\bibinfo
  {journal} {Phys. Rev. Lett.}\ }\textbf {\bibinfo {volume} {91}},\ \bibinfo
  {pages} {067003} (\bibinfo {year} {2003})}\BibitemShut {NoStop}%
\bibitem [{\citenamefont {Kubo}\ and\ \citenamefont {Toyabe}(1967)}]{Kubo1967}%
  \BibitemOpen
  \bibfield  {author} {\bibinfo {author} {\bibfnamefont {R.}~\bibnamefont
  {Kubo}}\ and\ \bibinfo {author} {\bibfnamefont {T.}~\bibnamefont {Toyabe}},\
  }\href@noop {} {\emph {\bibinfo {title} {Magnetic Resonance and
  Relaxation}}},\ edited by\ \bibinfo {editor} {\bibfnamefont {R.}~\bibnamefont
  {Blinc}}\ (\bibinfo  {publisher} {North-Holland},\ \bibinfo {address}
  {Amsterdam},\ \bibinfo {year} {1967})\BibitemShut {NoStop}%
\bibitem [{\citenamefont {Yaouanc}\ and\ \citenamefont
  {de~R\'eotier}(2011)}]{Yaouanc2011}%
  \BibitemOpen
  \bibfield  {author} {\bibinfo {author} {\bibfnamefont {A.}~\bibnamefont
  {Yaouanc}}\ and\ \bibinfo {author} {\bibfnamefont {P.~Dalmas}\ \bibnamefont
  {de~R\'eotier}},\ }\href@noop {} {\emph {\bibinfo {title} {Muon Spin
  Rotation, Relaxation, and Resonance: Applications to Condensed Matter}}}\
  (\bibinfo  {publisher} {Oxford University Press},\ \bibinfo {address}
  {Oxford},\ \bibinfo {year} {2011})\BibitemShut {NoStop}%
\bibitem [{\citenamefont {Aczel}\ \emph {et~al.}(2010)\citenamefont {Aczel},
  \citenamefont {Williams}, \citenamefont {Goko}, \citenamefont {Carlo},
  \citenamefont {Yu}, \citenamefont {Uemura}, \citenamefont {Klimczuk},
  \citenamefont {Thompson}, \citenamefont {Cava},\ and\ \citenamefont
  {Luke}}]{Acze2010}%
  \BibitemOpen
  \bibfield  {author} {\bibinfo {author} {\bibfnamefont {A.~A.}\ \bibnamefont
  {Aczel}}, \bibinfo {author} {\bibfnamefont {T.~J.}\ \bibnamefont {Williams}},
  \bibinfo {author} {\bibfnamefont {T.}~\bibnamefont {Goko}}, \bibinfo {author}
  {\bibfnamefont {J.~P.}\ \bibnamefont {Carlo}}, \bibinfo {author}
  {\bibfnamefont {W.}~\bibnamefont {Yu}}, \bibinfo {author} {\bibfnamefont
  {Y.~J.}\ \bibnamefont {Uemura}}, \bibinfo {author} {\bibfnamefont
  {T.}~\bibnamefont {Klimczuk}}, \bibinfo {author} {\bibfnamefont {J.~D.}\
  \bibnamefont {Thompson}}, \bibinfo {author} {\bibfnamefont {R.~J.}\
  \bibnamefont {Cava}}, \ and\ \bibinfo {author} {\bibfnamefont {G.~M.}\
  \bibnamefont {Luke}},\ }\bibfield  {title} {\enquote {\bibinfo {title} {Muon
  spin rotation/relaxation measurements of the noncentrosymmetric
  superconductor {Mg}$_{10}${Ir}$_{19}${B}$_{16}$},}\ }\href {\doibase
  10.1103/PhysRevB.82.024520} {\bibfield  {journal} {\bibinfo  {journal} {Phys.
  Rev. B}\ }\textbf {\bibinfo {volume} {82}},\ \bibinfo {pages} {024520}
  (\bibinfo {year} {2010})}\BibitemShut {NoStop}%
\bibitem [{\citenamefont {Biswas}\ \emph {et~al.}(2012)\citenamefont {Biswas},
  \citenamefont {Hillier}, \citenamefont {Lees},\ and\ \citenamefont
  {Paul}}]{Biswas2012}%
  \BibitemOpen
  \bibfield  {author} {\bibinfo {author} {\bibfnamefont {P.~K.}\ \bibnamefont
  {Biswas}}, \bibinfo {author} {\bibfnamefont {A.~D.}\ \bibnamefont {Hillier}},
  \bibinfo {author} {\bibfnamefont {M.~R.}\ \bibnamefont {Lees}}, \ and\
  \bibinfo {author} {\bibfnamefont {D.~McK.}\ \bibnamefont {Paul}},\ }\bibfield
   {title} {\enquote {\bibinfo {title} {Comparative study of the
  centrosymmetric and noncentrosymmetric superconducting phases of
  {Re}$_{3}${W} using muon spin spectroscopy and heat capacity measurements},}\
  }\href {\doibase 10.1103/PhysRevB.85.134505} {\bibfield  {journal} {\bibinfo
  {journal} {Phys. Rev. B}\ }\textbf {\bibinfo {volume} {85}},\ \bibinfo
  {pages} {134505} (\bibinfo {year} {2012})}\BibitemShut {NoStop}%
\bibitem [{\citenamefont {Wilson}\ \emph {et~al.}(2017)\citenamefont {Wilson},
  \citenamefont {Hallas}, \citenamefont {Cai}, \citenamefont {Guo},
  \citenamefont {Gong}, \citenamefont {Sankar}, \citenamefont {Chou},
  \citenamefont {Uemura},\ and\ \citenamefont {Luke}}]{Wilson2017}%
  \BibitemOpen
  \bibfield  {author} {\bibinfo {author} {\bibfnamefont {M.~N.}\ \bibnamefont
  {Wilson}}, \bibinfo {author} {\bibfnamefont {A.~M.}\ \bibnamefont {Hallas}},
  \bibinfo {author} {\bibfnamefont {Y.}~\bibnamefont {Cai}}, \bibinfo {author}
  {\bibfnamefont {S.}~\bibnamefont {Guo}}, \bibinfo {author} {\bibfnamefont
  {Z.}~\bibnamefont {Gong}}, \bibinfo {author} {\bibfnamefont {R.}~\bibnamefont
  {Sankar}}, \bibinfo {author} {\bibfnamefont {F.~C.}\ \bibnamefont {Chou}},
  \bibinfo {author} {\bibfnamefont {Y.~J.}\ \bibnamefont {Uemura}}, \ and\
  \bibinfo {author} {\bibfnamefont {G.~M.}\ \bibnamefont {Luke}},\ }\bibfield
  {title} {\enquote {\bibinfo {title} {$\mu${SR} study of the
  noncentrosymmetric superconductor {Pb}{Ta}{Se}$_{2}$},}\ }\href {\doibase
  10.1103/PhysRevB.95.224506} {\bibfield  {journal} {\bibinfo  {journal} {Phys.
  Rev. B}\ }\textbf {\bibinfo {volume} {95}},\ \bibinfo {pages} {224506}
  (\bibinfo {year} {2017})}\BibitemShut {NoStop}%
\bibitem [{\citenamefont {Agterberg}\ \emph {et~al.}(1999)\citenamefont
  {Agterberg}, \citenamefont {Barzykin},\ and\ \citenamefont
  {Gor'kov}}]{Agterberg1999}%
  \BibitemOpen
  \bibfield  {author} {\bibinfo {author} {\bibfnamefont {D.~F.}\ \bibnamefont
  {Agterberg}}, \bibinfo {author} {\bibfnamefont {V.}~\bibnamefont {Barzykin}},
  \ and\ \bibinfo {author} {\bibfnamefont {Lev~P.}\ \bibnamefont {Gor'kov}},\
  }\bibfield  {title} {\enquote {\bibinfo {title} {Conventional mechanisms for
  exotic superconductivity},}\ }\href {\doibase 10.1103/PhysRevB.60.14868}
  {\bibfield  {journal} {\bibinfo  {journal} {Phys. Rev. B}\ }\textbf {\bibinfo
  {volume} {60}},\ \bibinfo {pages} {14868} (\bibinfo {year}
  {1999})}\BibitemShut {NoStop}%
\end{thebibliography}%

\end{document}